# A joint analysis of the Drake equation and the Fermi paradox

Nikos Prantzos

Institut d'Astrophysique de Paris, UMR7095 CNRS, Université P. & M. Curie,
98bis Bd. Arago, 75104 Paris, France

prantzos@iap.fr

**ABSTRACT:** I propose a unified framework for a joint analysis of the Drake equation and the Fermi paradox, which enables a simultaneous, quantitative study of both of them. The analysis is based on a simplified form of the Drake equation and on a fairly simple scheme for the colonization of the Milky Way. It appears that for sufficiently long-lived civilizations, colonization of the Galaxy is the only reasonable option to gain knowledge about other life forms. This argument allows one to define a region in the parameter space of the Drake equation where the Fermi paradox definitely holds ("Strong Fermi paradox").

# 1. Introduction

Thinking on extraterrestrial life goes back at least as far as the Greek antiquity. The main argument for the existence of life forms and intelligent beings beyond Earth has been most clearly formulated by Metrodorus, disciple of Epicurus, in the 3d century B.C.: *"To consider the Earth as the only populated world in infinite space is as absurd as to assert that in an entire field sown with millet only one grain will grow."*. The idea of space being infinite, with an infinite number of atoms populating it and composing its various objects, was a key ingredient of the atomistic philosophy of Leucippus, Democritus and Epicurus.

The argument is also invoked today by proponents of extraterrestrial intelligence (ETI) essentially unaltered, although the concept of infinity is not used any more (because it is difficult to handle and it may lead to paradoxes, like e.g. in an infinite universe, everything – including ourselves – could exist in an infinite number of copies). It has been replaced by our Milky Way galaxy, which contains about 100 billion stars. That number is considered by some - mostly astronomers - to be large enough as to make Metrodorus' argument applicable to the Milky Way, while others – especially biologists – are not impressed by that number and remain skeptical concerning ETI.

In the second half of the 20$^{th}$ century, the debate on ETI was largely shaped by the Fermi paradox and the Drake equation. The latter was proposed in 1961 by American astronomer Frank Drake and became ever since the key quantitative tool to evaluate the probabilities for radio-contact with ETI (CETI), with the early evaluations being overly optimistic. The former was formulated in 1950 by Italian physicist Enrico Fermi, but remained virtually unknown until 1975, when it was resurrected by D. Viewing and independently re-discovered by M. Hart. In a concise form, it opposes a healthy skepticism to the optimistic views on ETI: *"(if there are many of them), where are they?"*

In this work, I propose a unified framework for a joint analysis of the Drake equation and the Fermi paradox, which enables a simultaneous, quantitative study on both of them. The analysis is based on a simplified form of the Drake equation and on a fairly simple scheme for the colonization of the Milky Way. It appears that for sufficiently long-lived civilizations, colonization of the Galaxy is the only reasonable option to gain knowledge about other life forms. This argument allows one to define a region in the parameter space of the Drake equation where the Fermi paradox holds strongly ("Strong Fermi paradox").

The plan of the paper is as follows: In Sec. 2, I analyze a simplified form of the Drake equation. In Sec. 3, I first provide some supplementary historical material on the Fermi paradox (Sec. 3.1), for which I propose an analytical formulation (Sec. 3.2). I then cast it in the same form as the simplified Drake equation by using a simple scheme for Galactic colonization (Sec. 3.3). Some of the implications are analyzed in Sec. 4.

## 2. The Drake equation

In its original formulation, the Drake equation reads

$N = R_* f_p n_e f_l f_i f_T L$     (1)

where $R_*$ is the rate of star formation in the Galaxy (i.e. number of stars per unit time), $f_p$ is the fraction of stars with planetary systems, $n_e$ is the average number of planets around each star, $f_l$ is the fraction of planets where life developed, $f_i$ is the fraction of planets where intelligent life developed and $f_T$ is the fraction of planets with technological civilizations. Obviously, *N* and *L* are intimately connected: if *N* is the number of radio-communicating civilizations – as in the original formulation by Drake – then *L* is *the average duration of the radio-communication phase* of such civilizations (and not their total lifetime, as sometimes incorrectly stated). On the other hand, anticipating on the content of Sec. 3, if *N* is meant to be the number of technological or space-faring civilizations then *L* represents the duration of the corresponding phase.

Although never explicitly stated (to my knowledge), the Drake formula obviously corresponds to the equilibrium solution of an equation similar to the well known equation of radioactivity $dN/dt=-N/L$, where *N* is the number of radioactive nuclei and *L* is their lifetime. In the steady state, where the production rate *P* is equal to the decay rate $D = dN/dt = -N/L$, one has $N = P L$. In a similar vein, the product of all the terms of the Drake formula except *L* can be interpreted as the production rate *P* of radio-communicating (or technological or space-faring) civilizations in the Galaxy.

It should be noticed that time does not appear explicitly in the terms of the Drake equation. This may be problematic, since we are interested in the present-day number of civilizations $N(t_0)$, while the stars harboring such civilizations were formed at time $t_0-T$, where T is probably a substantial fraction of the age of the Galaxy (T~4.5 Gyr in the case of our civilization), i.e. at a time where the rate of star formation $R_*(t_0-T)$ was, perhaps, very different from the present day star formation rate $R_*(t_0)$. However, in the case of the Milky Way, which is a Sbc-type spiral galaxy, there is convincing evidence for a "quiescent" evolution at quasi-constant pace over billions of years (Kennicutt and Evans 2012). In that case, the average star formation rate <R> over the Galaxy age A~10 Gyr can be considered as a fairly good approximation of R*(t) at any time t and, consequently, the equilibrium solution approximates well the real situation. Notice that the solution N<1 is acceptable in that case, meaning that the interval Δt between the occurrence of two such civilizations in the Galaxy is larger than their average duration L; in other terms, there is just one civilization in the Galaxy emerging at time t and lasting for a duration L, but the next one emerges at t+Δt where Δt>L. On the other hand, if $R_*(t)$ varies widely in time, like e.g. in elliptical galaxies – which formed practically all their stars in the first couple of Gyr – then the equilibrium solution does not apply and instead of the Drake formula one should use simply $N(t)=N_* \exp(-(t-T)/L)$, where $N_*$ is the total number of stars in that galaxy, formed in the first couple of Gyr.

In the fifty years since the formulation of the Drake equation, a fairly popular game consisted in attributing "plausible" numerical values to its various terms to estimate $N$. Unsurprisingly, estimates of different authors varied by many orders of magnitude: A. C. Cameron (1963) finds N=2 $10^6$ civilizations, almost the same number as the N=$10^6$ civilizations obtained by Sagan (1963) or Shklofski and Sagan (1966), while von Hoerner (1962) and Jones (1981) find N<100. It is significant to notice that average values in the early years were substantially larger than found in later times. In some cases, additional terms were added to the original ones to account for new astronomical factors (e.g. Ksanfomality 2004) or for intermediate steps in the development of a civilization, and even statistical treatments were considered, to account for dispersion in the values of the relevant parameters (e.g. Maccone 2010 or Glade et al. 2011).

Despite the apparent sophistication of such efforts, the game is rather meaningless, because until a few years ago only $R_*(t_0)$ could be estimated from observations. Only in the past few years statistics to estimate the second and third terms became available. The next four terms will remain unknown until we have a good theory on the emergence of life, intelligence and technology, or even better until such phenomena are detected beyond Earth (but then very few will care about the Drake equation anymore, since the goal will have been achieved). However, despite its inability to help calculate $N$ – which can have, in principle, any value between 1 and, say, $10^8$ - the Drake equation has been extremely useful, as it provided a framework allowing us to formulate our knowledge/thoughts/educated guesses about a very complex phenomenon such as the development of life and intelligence in the astrophysical setting of the Milky Way.

In this work, we shall follow a different approach from most previous ones: instead of introducing additional terms in Eq. (1), we shall condense its seven terms to only three. Our aim is twofold: 1) to illustrate quantitatively some implications of the number $N$ for SETI and CETI, and 2) to use exactly the same framework for a quantitative assessment of the Fermi paradox.

For that purpose, we rewrite the Drake equation as

$N=R_{ASTRO}f_{BIOTEC}L$     (2)

where $R_{ASTRO} = R_* f_P n_e$ represents the production rate of habitable planets (determined through astrophysics) and $f_{BIOTEC} = f_l \; f_i \; f_T$ represents the product of all chemical, biological and sociological factors leading to the development of a technological civilization. Obviously, $f_{BIOTEC} \leq 1$; its maximum possible value $f_{BIOTEC} = 1$ requires $f_l = f_i = f_T =1$ (a rather implausibly optimistic combination) but there is no constrain on its lower value.

The astrophysical factor $R_{ASTRO}$ is expected to be reasonably constrained in the foreseeable future. Indeed, its first term, $R_*$, is already constrained by observations in the Milky Way to be ~4 stars/yr (Chomiuk and Povich 2011 give ~1.9 $M_\odot$/yr for the present day star formation rate and there are ~2 stars per $M_\odot$ in a normal stellar initial mass function like the one of Kroupa 2003). However, its average past value was probably higher by a factor of 2-3, and we shall adopt the value of 5 $M_\odot$/yr which corresponds to an average star production rate of $R^* = <R> \sim 10$ stars yr$^{-1}$; this average SF rate reproduces well the stellar mass of 5 $10^{10}$ $M_\odot$ or the $10^{11}$ stars of the Milky Way if assumed to hold for the age of the Galaxy A~10 Gyr. Only 10% of those stars are appropriate for harboring habitable planets, because their mass has to be smaller than 1.1 $M_\odot$, i.e. they have to be sufficiently long-lived (with main sequence lifetimes larger than 4.5 Gyr) and larger than 0.7 $M_\odot$, to possess circumstellar habitable zones outside the "tidally locked region" (see Selsis et al. 2007). On the other hand, according to Mayor et al. (2011), the statistics currently available on extra-solar planets suggest that about 13% of the surveyed stars possess super-Earths, i.e. planets in the 3-30 $M_\oplus$ range. We shall consider here that this fraction describes the product $f_p n_e$ in the Drake equation, corresponding to *stars with continuously habitable planets* (i.e. Earth-like planets orbiting continuously their star within the circumstellar habitable zone). This is a fairly optimistic estimate, its only merit being that it imposes a plausible upper limit on the fraction of such stars. Combined with the aforementioned formation rate of 0.7-1.1 $M_\odot$ stars, it leads to $R_{ASTRO}$ =0.1 habitable planet per year. We shall adopt this value here and we shall investigate the space of the remaining parameters $f_{BIOTEC}$ and *L,* which are totally unknown at present.

We assume that, to a first approximation, we can describe the Galactic disk by a cylinder of radius $R_G$ =12 kpc and height h=1 kpc, where the *N* civilizations of the Drake equation are distributed uniformly. By equating the volume of the Galactic cylinder V=π $R_G^2$h with the sum of *N* volumes of spheres of average radius r occupied by each civilization (Fig. 1), one obtains the average distance between two civilizations as D = 2r = 2 $\left(\frac{3V}{4\pi N}\right)^{1/3}$ for the case where D<h. In the case of a small number of civilizations (say *N*<1000) it turns out that D>h and a more appropriate expression is then D = 2r = 2 $R_G/\sqrt{N}$.

The values of the average distances D as a function of the number of civilizations *N* appear in Fig. 2. For less than a thousand civilizations typical distances are larger than 3000 light-years while for 10 million civilizations they are of the order of 100 light-years. Notice that *N* is the number of civilizations co-existing in the Galaxy during *L.* For each number *N* there is a minimum value $L_{MIN}$, corresponding to $f_{BIOTEC}$=1 in Eq. (2) for the adopted value of $R_{ASTRO}$=0.1 per yr; the corresponding curve also appears in Fig. 2. Obviously, communication between neighboring civilizations requires their duration *L* to be larger than twice the travel-time D/c (where c is the light speed) of radio-waves. An inspection of Fig. 2 shows that if there are less than a few hundred co-existing civilizations in the Galaxy, their radio-emission phase has to last longer than $10^4$ years to allow them to establish radio-communication.

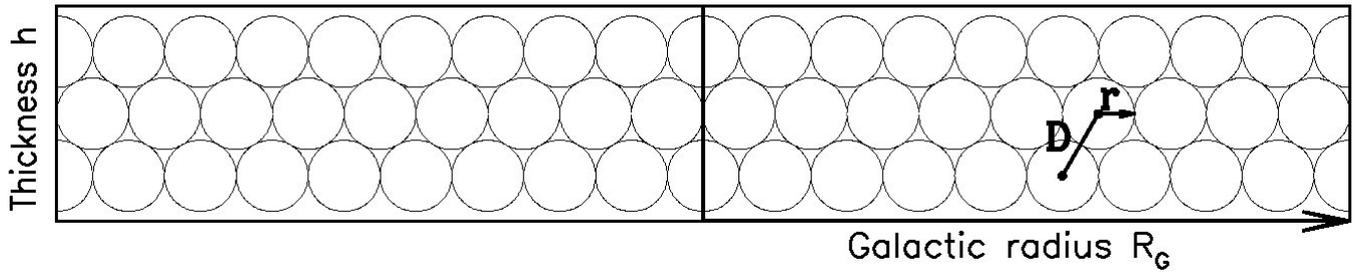

**Fig. 1** Filling the Galaxy with N civilizations located at average distances D of each other.

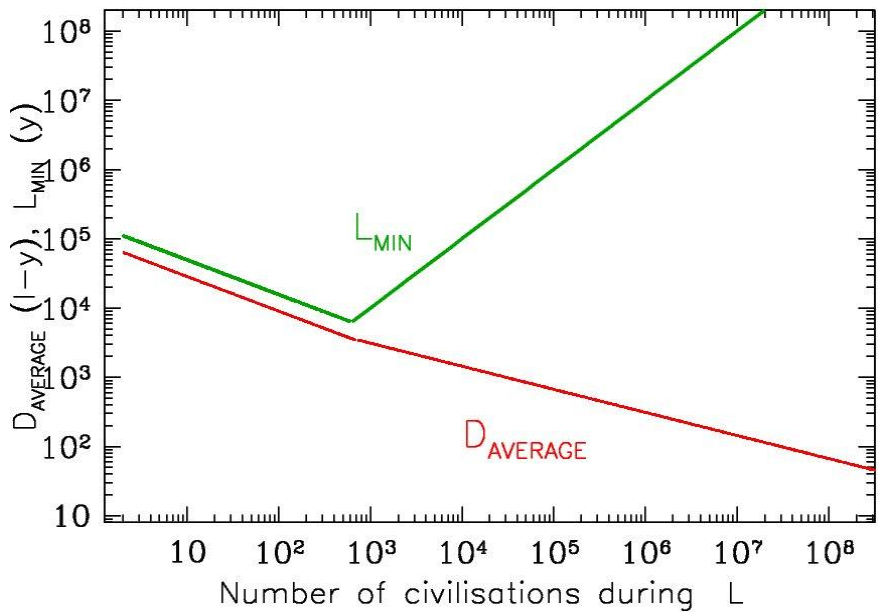

**Fig. 2** Average distances D (continuous curve) between civilizations as a function of their total number N in the Galaxy, assuming they are uniformly distributed in the Milky Way disk as in Fig. 1. The upper curve $L_{MIN}$ fulfills simultaneously the conditions L=N/0.1 (for $R_{ASTRO}$=0.1 and $f_{BIOTEC}$=1) and L=2D/c (for two-way communication between neighboring civilizations).

Notice that these numbers assume that ALL civilizations have similar values of L, i.e. that dispersion *ΔL* in *L* is much smaller than *L* itself. This need not be the case. Statistical treatments, considering *ΔL~L* and canonical distributions have been recently applied by Maccone (2010b) and Glade et al. (2012). However, in any case, the unknown mean value of *L* plays a more important role than the equally unknown form of its distribution; we shall focus then here only on that mean value *L*, leaving the implications of a statistical treatment for a future work.

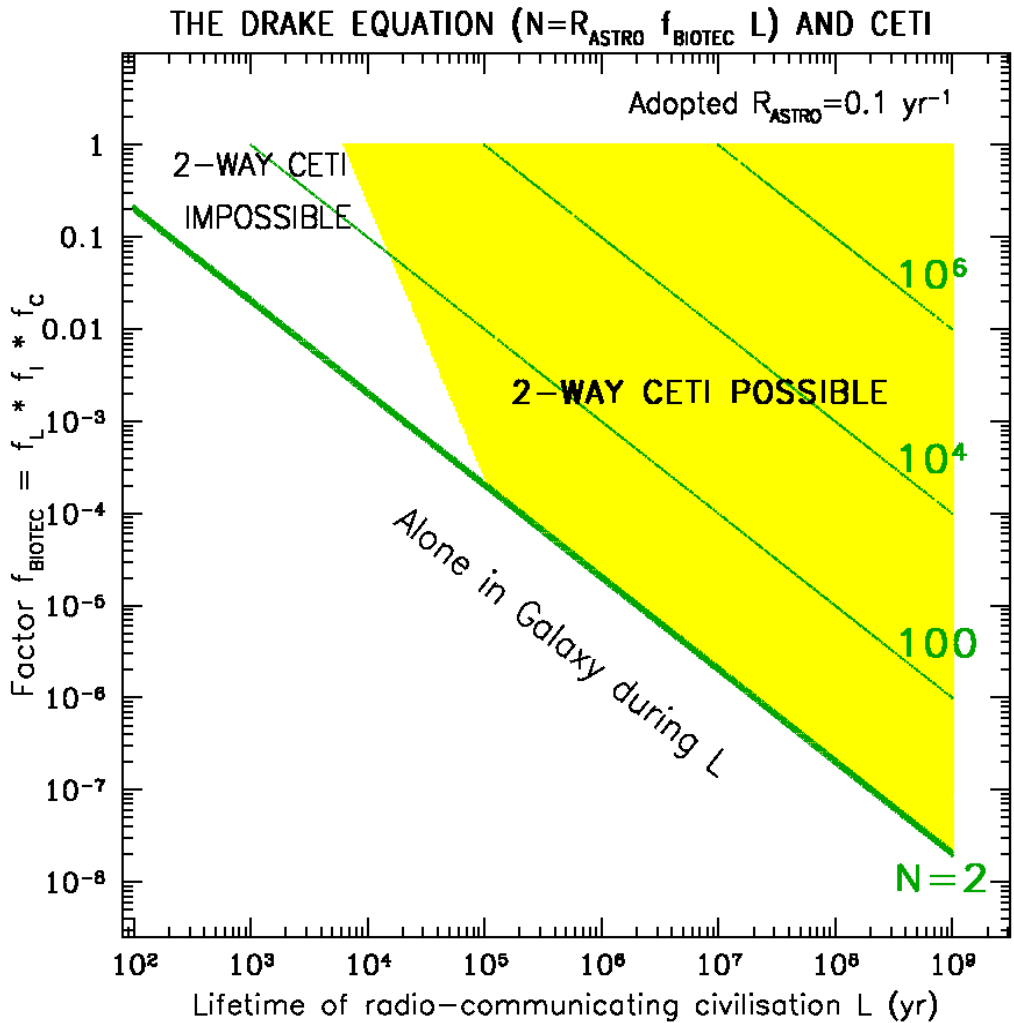

**Fig. 3:** The parameter space of $f_{BIOTEC} = f_L f_I f_C$ (=bio-sociological factor) vs L (the average lifetime of a technological civilization). Calculations are done by assuming a value $R_{ASTRO}=R_* f_P n_e =0.1$ habitable planet per yr for the astronomical factor (see text) in the Drake equation. This leads to values for the number N of technological civilizations indicated by the parallel diagonal lines. N=2 is the minimum required for two-way communication during L. The shaded region allows for two-way communication between two civilizations during L, i.e. it satisfies the condition L>2D/c, where D is the average distance between civilizations (see Figs. 1 and 2) and c the velocity of light.

For the adopted value of $R_{ASTRO}$, a given value of $N$ (and thus a corresponding value of D) represents a line in the $f_{BIOTEC}$ vs $L$ diagram (Fig. 3), with $N$=2 being a lower limit for a communication. The time required for such a communication, involving an exchange of radio-signals, is T=2D/c. The condition for radio-communication ( T<$L$) bounds the $f_{BIOTEC}$ vs $L$ diagram from the left, according to

$N > 48\ R_G^2 h/L^3$    for D < h, and

$N > 16\ R_G^2$         for D > h

In those expressions, distances are expressed in light-years and durations in years, allowing thus one to omit the light velocity (c=1).

An inspection of Fig. 3 shows that, for the adopted value of $R_{ASTRO}$=0.1 habitable planets per year (an upper limit), communications are possible only for civilizations spending at least 10000 years in the radio-communication phase. If $f_{BIOTEC}$=1, there are about 1000 such civilizations in the Galaxy and our chances of eavesdropping the communications of our closest neighbours are not negligible. But if $f_{BIOTEC}$=$10^{-3}$, there are just a few civilizations in the Galaxy and chances of eavesdropping appear insignificant. Of course, the larger the duration of the radio-communication phase, the larger is *N* (for the same value of $f_{BIOTEC}$) and chances improve considerably. However, it should be noticed that there is a large region of the parameter space (for all *L*<10000 yr, irrespectively of $f_{BIOTEC}$) for which communications are impossible or there is no second radio-emitting civilization in the Galaxy during *L*.

One may conclude then that it is improbable that we communicate with (or eavesdrop) civilizations of a level either similar to ours (L~100 yr of radio-emission) or even radio-emitting for a few thousands of years. Only civilizations emitting for much longer timescales have chances to be detected by our SETI programs. On the other hand, it may well be that in the case of $f_{BIOTEC}$<$10^{-2}$ we may spend thousands of years - and in the case of $f_{BIOTEC}$<$10^{-5}$ even millions of years - searching for radio-signals with no effect, since we would be alone in the Galaxy during that period. As already stressed, those numbers correspond to the optimistic case of $R_{ASTRO}$=0.1 habitable planet per year. It seems obvious that such considerations will affect the strategy of any extraterrestrial civilizations searching their siblings.

At this point it should be noticed that it is a rather futile exercise to try to imagine the kind of communications that technological civilizations older than a few centuries might have. In particular, the problem of understanding alien messages has been given serious attention by soviet astronomers in the 60ies and 70ies (as properly emphasized in the recent monograph by Sheridan 2009) but it appears to be virtually absent from the recent literature on the subject.

**3. The Fermi Paradox**

In this section we reassess the Fermi paradox in the context of the parameters of the Drake equation, as discussed in Sec. 2. The origin of this famous paradox is documented by Eric Jones in his 1985 Los Alamos report: it appears that Fermi formulated his question "Where is everybody?" during a lunch time conversation at Los Alamos in the summer of 1950 with colleagues Emil Konopinsky, Herbert York and Edward Teller. Its early history is related by Stephen Webb in his book "*Where is everybody?*", perhaps the most complete investigation today of the solutions proposed to solve the paradox. Webb provides a detailed account of the early ideas on the paradox, including its earlier discovery by Russian father of astronautics K. Tsiolkovski in 1933, its independent re-discovery by M. Hart in 1975 and its identification as a paradox by D. Viewing in 1975. In recognition of those early contributions to the debate, Webb calls it the "Tsiolkovsky-Fermi-Hart-Viewing" paradox. We shall keep the shorter term "Fermi paradox" here for convenience.

In this section, we first provide some supplementary material on the pre-history and early history of the Fermi paradox. We then present it in a more formal way, making explicit the underlying assumptions. Finally we analyze it in the same way as we did for CETI in the previous section, i.e. in terms of the factors $f_{BIOTEC}$ and $L$ of the Drake equation.

### 3.1 On the prehistory and early history of the paradox

The basic idea of the ``paradox'' was already formulated more than three centuries ago, albeit in an inconclusive way. In 1686, the French novelist Bernard le Bovier de Fontenelle (who later became secretary of the French Academy of Sciences) published his best selling book *Entretiens sur la Pluralité des Mondes* (*Conversations on the Plurality of Worlds*). It is often considered as the first popular science book and it is written in the form of dialogues between the author and a charming and ingenuous marchioness. The marchioness counters the author's assertion that ``intelligent beings exist in other worlds, for instance the Moon'' by the retort: ``If this were the case, the Moon's inhabitants would already have come to us before now''. Fontenelle can only argue that the time to master space travel is probably too long and that the Moon's inhabitants ``at this time are maybe exercising themselves; when they shall be more able and more experienced, we may see them... after all, the Europeans did not arrive in America till nearly at the end of six thousand years'' (in Fontenelle's days, the Universe was thought to be 6000 years old, based on biblical accounts). Note that Fontenelle's time argument is completely symmetric between Earth's and Moon's inhabitants, namely they both have at most 6000 years to master space travel, but there is no hint as to whether one of the two civilizations is more advanced than the other; thus, the absence of people from the Moon on Earth could not really be considered as a surprise, since earthlings had not visited the Moon either. The modern argument by Fermi explicitly assumes, by virtue of the Copernican principle, that some of the extraterrestrial civilizations are (considerably) older than ours and therefore they have had time enough to spread into the Galaxy and reach our planet, while our own civilization is unable to do so at present.

For more than ten years after the now famous lunch-time discussion at Los Alamos, there appears to be no written trace of Fermi's question (at least to my knowledge). The earliest trace I am aware of is a footnote in a paper published by American astronomer C. Sagan in 1963: "This possibility has been seriously raised before; for example, by Enrico Fermi, in a now rather well known dinner table discussion at Los Alamos during the Second World War, when he introduced the problem with the words *Where are they?*". Sagan provides no information on his sources, making it difficult to know where he got such erroneous information about a "dinner table discussion during Second World War". A few years later, the phrase "*Where are they?,* attributed to Enrico Fermi but without any comment, appeared in the book of I. Shklofski and C. Sagan "Intelligent Life in the Universe". Obviously, Sagan realized that the question might have some profound implications for the CETI endeavor, but he could not or did not wish to brood over them.

*3.2 An explicit formulation*

Any paradox is based on at least one invalid assumption. The logical statement of Fermi's paradox runs as follows (see also Prantzos 2000):

A. Our civilization is not the only technological civilization in the Galaxy.

B. Our civilization is in every way ``average'', or typical. In particular, it is not the first to have appeared in the Galaxy, it is not the most technologically advanced, and it is not the only one seeking to explore the cosmos and communicate with other civilizations.

C. Interstellar travel, although impossible for us today, is not too difficult for civilizations slightly more advanced than ours. Some extraterrestrial civilizations have mastered this kind of travel and undertaken a galactic colonization program, either with or without self-replicating robots (F. Tipler 1980).

D. Galactic colonization is a relatively fast undertaking and could be achieved in a relatively small fraction of Galaxy's age (less than a few hundred million years).

If hypotheses A to D are valid, one clearly deduces that ``they should be here''. Supporters of ETI reject at least one of assumptions C and D, and some even go so far as to deny B, in order to save their key hypothesis A. In contrast, their opponents uphold the plausibility of C and D, whilst completely rejecting B. By virtue of the Copernican principle, hypothesis A should then be rejected as well. Of course, if the Copernican principle is inapplicable in case B, hypothesis A can still be saved (but at what a price!).

Most of the arguments in the debate on the Fermi paradox are of sociological nature (see Webb 2002 for a thorough, albeit not exhaustive, census) and concern assumptions B and D above. Here I will assume that assumptions A to C are valid and I will explore the consequences of D, in the framework of the Drake equation (Eq. 2), as discussed in Sec. 2.

*3.3 The Fermi Paradox in terms of the Drake equation*

The Fermi paradox has been quantitatively explored by various techniques using population dynamics (Viewing 1975, Hart 1975, Newman and Sagan 1981 etc.), percolation analysis (Landis 1998) or Monte Carlo methods (e.g. Jones, 1981, Bjoerk 2011). Depending on the underlying assumptions, quite different results are found on the timescales of Galactic colonization. In most cases, timescales are found to be much smaller than the age of the Galaxy (e.g. Jones 1981), thus substantianting the Fermi paradox, while in a few cases much larger values are found (e.g. Bjoerk 2011).In fact, Wiley (2012) shows that none of the assumptions underlying the aforementioned models is strong enough to render the conclusions robust.

Here I illustrate the Fermi paradox in a fairly simple way, by casting it into the same form as the Drake equation in Fig. 3. I assume that once a civilization masters the techniques of interstellar travel, it starts a thorough colonization/exploration of its neighborhood for a period *L*. Colonization proceeds in a directed way, i.e. it concerns only stars harboring nearby habitable planets, which are detected before the launching of the spaceships. According to the discussion in Sec. 2, an upper limit to the total number of such "interesting stars" in the Galaxy (i.e. by assuming $R_{ASTRO}$=0.1 per yr) is $n_i \sim 10^9$ stars, one hundredth of the total number of Galactic stars; their average density is then $\rho \sim 6 \cdot 10^{-5}$ stars per l-yr$^3$ and their average distances d~31 l-yr.

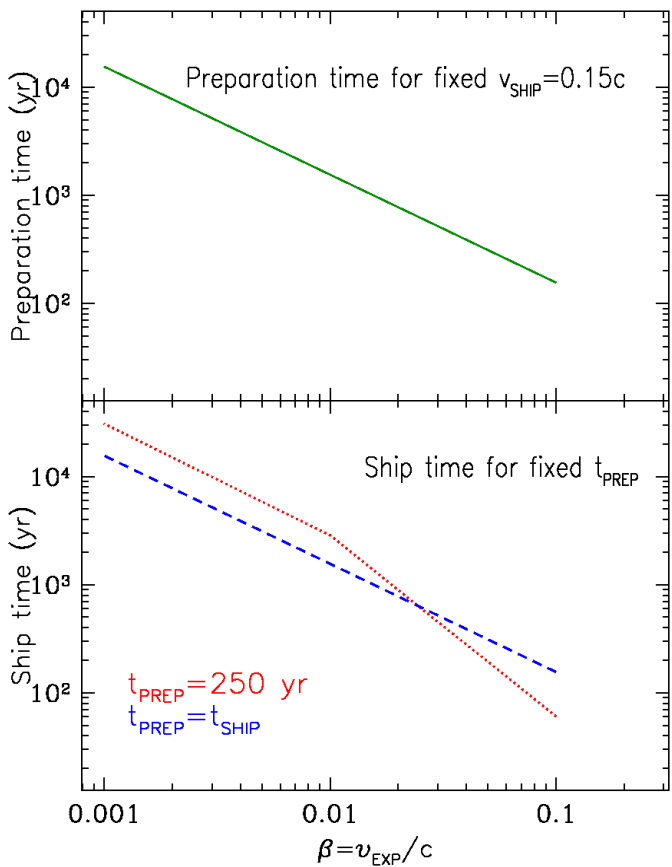

**Fig. 4:** Preparation time for new missions for fixed ship velocity (upper panel) and ship time for fixed preparation time (lower panel).

The colonization front expands outwards at an average velocity v=βc. Ships are sent to new stars not from the mother planet but from the colonized planets in the wavefront and they are launched after a time interval $t_{PREP}$ following colony foundation. This gives enough time to the colonizers to install on the new planet and prepare the next colonizing mission. Notice that v is the *effective velocity of the colonization front* and not the velocity of the interstellar ships; the latter is given by $v_{SHIP}=d/(d/v-t_{PREP})$ and has to be larger than v. Fig. 4 illustrates two different strategies of colonization in that framework. In the upper panel, the ship velocity is fixed to $v_{SHIP}$=0.15 c, thus the time spent in the ship to cross the average interstellar distance d is $t_{SHIP}=d/v_{SHIP}$ =206 yr and the preparation times for the new missions are $t_{PREP}$=31000, 2900, and 103 yr for expansion velocities v/c=0.001, 0.01 and 0.1, respectively. The bottom panel corresponds to the case where the preparation time is fixed either to $t_{PREP}$=250 yr or to be equal to the time spent in the ship $t_{SHIP}$; in both cases, ship times are found to be approximately $t_{SHIP}$~d/v, i.e. they are quite large for low v values. Only generation starships could be used for such an endeavor (unless the lifetimes of the extraterrestrials are at least as long as $t_{SHIP}$).

In the upper panel of Fig. 5 we display the radius r=vt of the colonization front as a function of time for the same three values of the velocity β=0.1, 0.01 and 0.001, respectively. In the middle panel appears the number of "interesting stars" engulfed within the colonization front after time t, i.e. $n(t)=\rho\ 4\pi/3\ (vt)^3$. The rate of new stars explored per unit time $dn/dt= \rho\ v\ 4\pi r^2$ is shown in the bottom panel of Fig. 5. For β=0.1, a few thousands of "interesting stars" are explored per year towards the end of the Galactic colonization, while for β=0.001 only a few stars are explored per year. In all cases, the exploration rate is much smaller at earlier times, while the average rate of colonized planets is obviously $n_i/L$. The rate of new stars per year appears large, but it should be recalled that they are supposed to be visited by probes launched by all the stars discovered in the previous time-step; the number of the latter is slightly smaller than the one of the former, so that only 1.1-1.2 ships per colonized planet are required to be launched further out, as to maintain constant the velocity of the colonization front. Two centuries are certainly enough for each new colony to prepare and launch a couple of new ships. We find then that full exploration of the Galaxy by a single civilization and its off-springs would take approximately ΔT=4 $10^5$, 4 $10^6$ and 4 $10^7$ years for values of β=0.1, 0.01 and 0.001, respectively (see Fig. 4). And even if some of the colonies within the expanding spheres die out, the spheres need not remain partially hollow for a long time, since (some of) the active colonies could send new ships back and revive the dying colonies.

The model adopted here for the expansion of a civilization in the Galaxy corresponds to the so-called "*Coral* model*", describing how corals grow in the sea. This model is presented, for instance, in the book « *Life in the Universe* » (Benett and Shostak 2007), and has been used by e.g. Maccone (2010a) in an attempt to provide a statistical description of galactic colonization. In both cases, a factor k=1/2 is introduced in front of the fraction defining the ship speed to account for the zig-zag motion of the colonizing ships in space (sometimes colonizing back some stellar systems previously left behind or re-activating dying colonies). Introducing this factor should change slightly the numerical results presented here. Alternatively, the results should remain the same, under the assumption of a speed twice as large as the values we adopted.

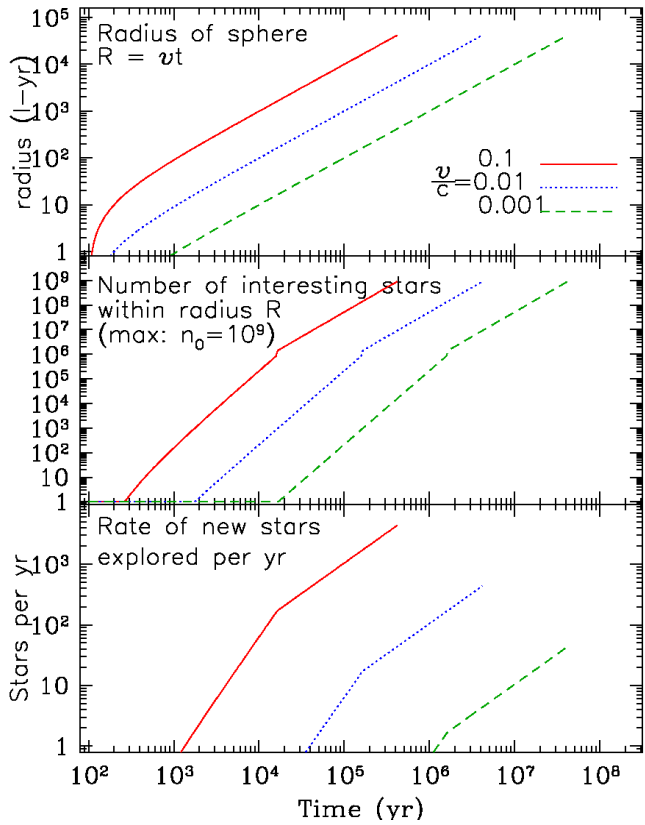

**Fig. 5:** Expansion of a space-faring civilization at constant speed v=βc, with β=0.1, 0.01 and 0.001 respectively. *Top panel*: Radius of the spherical expansion front as function of time. *Middle panel*: Number $n$ of "interesting stars" (i.e. 10% of the stars in the 0.7-1.1 $M_\odot$ range, assumed to harbor habitable planets) that are engulfed by the sphere of radius $r = v\,t$. *Bottom panel:* Corresponding rate $dn/dt$ of new stars explored per year for the same three values of β. In all panels, the curves stop when the total number $n_0=10^9$ of "interesting stars" is reached. Slopes of curves change when the radius $r$ of the expansion front becomes larger than the thickness $h$ of the Galactic disk.

The analysis of Fig. 5 concerns the colonization of the Galaxy in the case of a single colonizing civilization, starting from a single place in the Galaxy (a place close to the Galactic center was assumed here, but the results would not be very different if any other place was chosen). In Fig. 6 we extend the analysis to the case of N independent civilizations - as given by the Drake formula - in a similar form as the one adopted for the analysis of CETI in Fig. 3. We examine whether the Galactic volume can be filled within time $L$ by $N=R_{ASTRO}\,f_{BIOTEC}\,L$ spheres of radius $r=\beta cL$ for β=0.1, 0.01 and 0.001. If it can be filled, then every corner of the Galaxy should be visited by at least one of the N civilizations within its space-faring phase and the Fermi paradox holds for the corresponding values of $f_{BIOTEC}$ and $L$. Notice that here the lifetime $L$ corresponds to a single civilization and includes all its offspring colonies; in other terms, the colonies do not count as different civilizations, since they do not originate in an independent way.

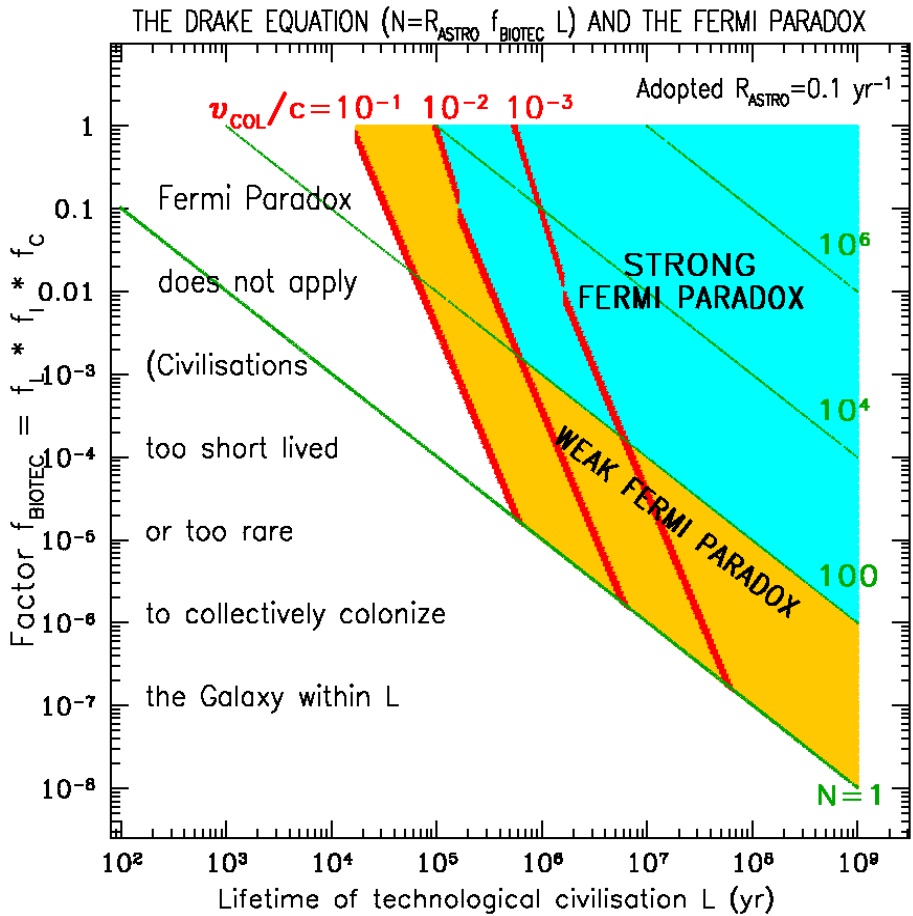

**Fig. 6:** The Fermi paradox presented in the $f_{BIOTEC}$ vs. L plane, in terms of the factors appearing in the Drake equation. In the shaded region, space-faring civilizations may collectively explore/colonize the whole Galaxy within their average duration L, assuming the colonization front expands with average velocity v/c=0.1, 0.01, and 0.001, respectively (thick lines, with slope changing at the point when the radius of the expansion front of each civilization becomes larger than the adopted height of the disk, i.e. for N~1000). The Fermi paradox applies in the shaded region. The region of the "Strong Fermi Paradox" is discussed in the text.

Results are displayed in Fig. 6. Depending on β and L, full colonization of the Galaxy by the N civilizations may or may not be possible during L. For instance, in the ultra-optimistic case of $f_{BIOTEC}$=1, a thousand civilizations would need only a couple of $10^4$ yr to collectively colonize the Galaxy at v/c=0.1, while a single civilization would need a hundred times longer. For values of $f_{BIOTEC}$ and L outside the shaded region, the Fermi paradox is no more a paradox: civilizations are too rare or too short lived to fully colonize the Galaxy within the duration L of their space exploring phase. Within the shaded region, the Fermi paradox holds, since such civilizations can, in principle, colonize the galaxy and they should have found us already.

In the case of a small number of civilizations, any one of the sociological ideas put forward to explain the Fermi paradox may be valid: indeed, some civilizations may never master space travel, or never wish to colonize or to perturb other, less mature, ones, or they may abandon their colonization effort shortly after they started it etc. (see Webb 2002). However, as emphasized in Prantzos (2000), such arguments appear hardly plausible in the case of a large number of independent civilizations. We shall assume then, somewhat arbitrarily, that for N>100 such sociological arguments can hold for some but not for the majority of civilizations. We shall also assume – in an equally arbitrary way - that an expansion velocity of v/c=0.01 for the colonization front is much more reasonable (less demanding) than v/c=0.1. Those assumptions define a region in the $f_{BIOTEC}$ vs. L plane (bounded from below by N=100 and from the left by v/c=0.01) where the Fermi paradox is arguably on a solid basis ("strong Fermi paradox") whereas the situation is less clear for the remaining shaded region ("weak Fermi paradox").

Ostriker and Turner (1986) argued that, even if advanced technological civilizations are common, they are unlikely to fully occupy the Galaxy, because at some point of their expansion, their mutual interactions could reduce the pace of colonization, leaving some portions of the Galaxy unoccupied for periods of the order of *L*. They base their arguments on a mathematical analysis drawing from the ideas of theoretical ecology and they suggest that the Earth may be found in such an unoccupied region, thus providing an(other) explanation of the Fermi paradox. Somewhat counter-intuitively, they claim that the complexity of the system describing population expansion allows one to predict its behavior with a reasonably high degree of confidence. Unfortunately, at the current stage of our knowledge, it is impossible to know whether Galactic colonization is better described by simple models (such as the "Coral-model" adopted here) or by more sophisticated ones, such as theirs.

## 4. Discussion

The novelty of this study is twofold: a) the presentation of the Drake formula in a diagram of $f_{BIOTEC}$ vs *L* (for a given value of $R_{ASTRO}$) in Fig. 3, and b) the presentation of the Fermi paradox in exactly the same form (Fig. 6), allowing one to establish some quantitative conclusions.

An inspection of Figs. 3 and 6 shows that the shaded regions (allowing CETI and full colonization of the Galaxy, respectively) are of similar size, i.e. they correspond to the same range of values of $f_{BIOTEC}$ and L. This occurs despite the fact that communication velocities are c, while maximum velocities of colonization fronts are v=0.1 c (in the most optimistic case). The reason is that CETI involves at least two signals exchanged over distances D=2r (see Fig. 1), i.e. a total distance of 4r crossed at light velocity c, which takes just 40% of the time required by the colonization front to cross radius r at velocity v=0.1 c. This difference by a factor of ~2 between the effective velocities of the communication carriers (electromagnetic signals vs. starships) is barely visible in the shaded regions of figures 3 and 6.

Although sending and receiving radio-signals is certainly a much easier enterprise than launching starships, the latter should not be too difficult either for a >$10^4$ yr old technological civilization (minimum age to have reasonable chances for CETI according to Fig. 3). Assuming that such civilizations wish to learn about other life forms (intelligent or not), one may ask what kind of action they might adopt: just content themselves with a program of radio-signal emission/reception, or undertake a serious effort of interstellar colonization? The former, although less demanding, might provide no results for several more $10^4$ yr (and even for millions of years, in case there are just a few such civilizations in the Galaxy at that period). The latter, even at the slow pace of 0.01 c, would bring within their grasp, in that same period, tens of thousands of stars, harbouring perhaps interesting life forms, or even civilizations as advanced as the Egyptians or the Greeks, yet unable to radio-communicate. Obviously, the benefits would be far greater in the latter case, pushing most (if not all) technological civilizations older than a few $10^3$ yr to undertake a serious program of interstellar exploration. And even if some of them abandon the program for various reasons, the effort of their neighbors could compensate for that.

In summary, it appears that although radio-communications constitute a natural means for SETI for civilizations younger than a few millennia, older civilizations should rather develop extensive programs of interstellar colonization, because this is the only way to achieve undisputable evidence (either for or against the existence of ETI) within their lifetime L. In those conditions, the Fermi paradox appears all the more paradoxical: if, as the SETI proponents claim, there are literary thousands of such advanced civilizations wishing to establish contact, "where are they?"

Today the "Plurality of worlds", as the field was known in the Antiquity, is more controversial than ever. Arguments on both sides (``It is unlikely that we are alone, in view of the Copernican principle and of such a large number of stars in the Galaxy'' and ``If there are so many of them, where are they?'') are of statistical kind. They are consequently of little import, for statistics cannot be based on the single case provided by life on Earth. Detection of life signatures on another planet would be a powerful reason to undertake interstellar travel, at first by sending unmanned probes. Detection of some extraterrestrial civilization would undoubtedly be one of the major landmarks in the history of mankind. On the other hand, non-detection of ETI signals, even after millennia of research, would never prove that there were no extraterrestrial civilizations. But it would be reason to prepare ourselves for a life of cosmic solitude.

**Acknowledgements:** I am grateful to two anonymous referees for useful comments which improved the text.